\documentstyle[aps,epsfig,floats]{revtex}

\begin{document}

\twocolumn[\hsize\textwidth\columnwidth\hsize\csname@twocolumnfalse%
\endcsname

\title{Disorder induced roughening transition of
many elastic lines in a periodic potential}
       
\author{T. Knetter$^1$, G. Schr\"oder$^1$, M. J. Alava$^{2}$,
and H. Rieger$^{3}$}

\address{$^1$ Institut f\"ur Theoretische Physik, Universit\"at zu
  K\"oln, 50937 K\"oln, Germany}
\address{$^2$ Helsinki University of Technology, Laboratory of
Physics, P.O.Box 1100, 02015 HUT, Finland}
\address{$^3$ Theoretische Physik, Universit\"at des Saarlandes,
66041 Saarbr\"ucken, Germany}

\date{\today}

\maketitle

\begin{abstract}
  The competing effect of a periodic pinning potential and random
  point disorder is studied for arrays of elastic lines or directed
  polymers. The groundstates are investigated by exact combinatorial
  optimization. In both two and three dimensions a phase
  diagram is found with two or three distinct phases: a strictly flat
  phase if the disorder is bounded and weak, a weakly
  fluctuating phase for intermediate valleys depths, where the lines
  roughen individually on a scale smaller than the line-line distance,
  and a rough phase for strong disorder, where
  the roughness follows the scaling with pure point disorder. The line
  wandering in the transverse direction leads in the
  three-dimensional, rough phase to an entangled state with a
  complicated topology.
\end{abstract}


\pacs{74.60.Ge 64.70.Rh 68.35.Ct 68.35.Rh}

]

Ensembles of elastic lines form one of the most interesting examples
of the interplay of disorder and an ordering tendency.  Much of the
attention is motivated by the experimental connections to dirty
type-II superconductors, in which flux lines (FL) interact with point or
columnar defects \cite{Blatter,Natt,Young}.  This interaction gives
often rise to beneficial, technologically important effects since the
impurities pin lines, individually or collectively and creating them
in a controlled fashion has been demonstrated by various techniques.
The theory of arrays of FLs with a wide variety of possible
disorder backgrounds has received lots of interest \cite{flls}.  It is
worth noting that the physics of an individual FL has
connections to paradigmatic questions in non-equilibrium statistical
mechanics \cite{halpin,laessig} through mappings to the
Kardar-Parisi-Zhang equation of kinetic roughening in surfaces, and to
the Burgers' equation of vortex-free turbulence \cite{kardar}.

Here we analyze with exact numerical tools the roughening or
disordering of arrays of elastic lines at zero temperature when point
disorder competes with a periodic potential that tries to order the
array into a regular structure. The FL analogy is an Abrikosov lattice
in the presence of point disorder. For both two- and three-dimensional
(2D and 3D) arrays of lines we find a transitions in terms of the
ratio $q = \left<\epsilon\right>/\Delta$, where $\Delta$ measures the
depth of the potential valleys, and $\left<\epsilon\right>$ is the
average strength of the point disorder. 
%
%
The picture we gain from our study is in contrast to recent renormalization
group studies of a model for {\it elastic periodic} media with point disorder
in the continuum limit \cite{Bouchaud92,emig}. Based on these one would expect
for an arbitrary disorder strength $q$ in two dimensions asymptotically (in
system size) a rough state with diverging FL displacements and line-to-line
displacement correlations, whereas we find this to be true only above
$q_{c_2}$. In other words, the lattice model for a flux line array with a
periodic potential shows a roughening transition which should, at least in 2d,
be absent according to analytical theories for an apparently related continuum
model. This is also surprising since for the same system without periodic
potential good agreement between analytical \cite{sos_ana} and numerical
\cite{sos_num} results has been achieved. After presenting the numerics
we will discuss the apparent contradiction with analytical results.

We are interested in an assembly of $N$ elastic lines
described by the Hamiltonian
\begin{eqnarray}
{\mathcal H} = &{\displaystyle\sum_{i=1}^N} 
{\displaystyle\int_0^H dz}
\Bigl\{ {\gamma\over2}\left[\frac{d{\bf r}_i}{dz}\right]^2
+\sum_{j(\ne i)}V_{\rm int}[{\bf r}_i(z)-{\bf r}_j(z)]
\nonumber\\
&+V_r[{\bf r}_i(z),z] + V_p[{\bf r}_i(z)]\Bigr\}\;.
\label{cont}
\end{eqnarray}
${\bf r}_i(z)\in {\cal R}^{d-1}$ (with $d=2$ and $d=3$ studied here)
is a displament vector, while $z$ is the longitudinal coordinate in a
system of height $H$; $V_r[{\bf r},z]$ describes the point disorder,
which we take to be delta-correlated with variance $\epsilon$; $V_{\rm
  int}[{\bf r}-{\bf r}']$ is a short-range repulsive interaction
between the lines (e.g.\ hard-core) and $V_p[{\bf r}]$ a periodic
potential with period $a$ in all transverse space directions.
We assume its minima to be well localized, i.e.\ they
have a width that is small against the interaction
range of the lines. This allows only single occupancy of the potential
valleys and we concentrate on the case in which the line density
($\rho=L^{d-1}/N$) is such that each potential valley is occupied by
exactly one elastic line, i.e.\ $\rho=1/a^{d-1}$. A similar model has
been studied in $d=1+1$ in \cite{HNV} for the competition between
point disorder and random columnar defects, i.e.\ a potential $V_p$
that is not periodic but random.

\begin{figure}[t]
\epsfxsize=\columnwidth\epsfbox{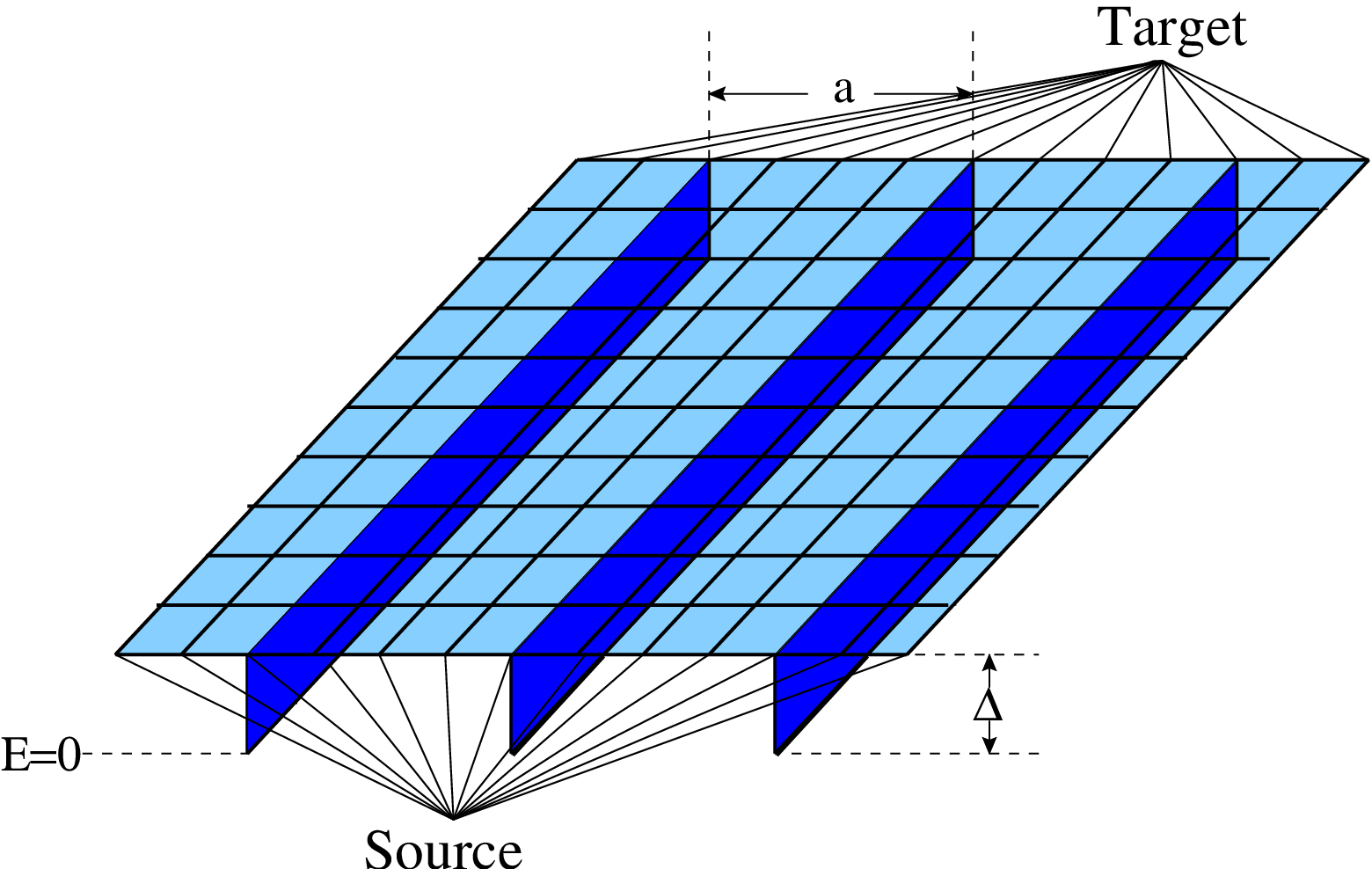}
\caption{
 Periodic potential in $2D$. The depth of the valleys is denoted
 by $\Delta$ and the nearest neighbour distance by $a$. Additional
 point disorder $\epsilon$ accomplishes the energy landscape. The FL
 can only enter and leave the system via the energetically neutral
 arcs connecting the source respectively the sink with the potential valleys.
 \label{hamfig}
}
\end{figure}

We study a lattice version of the continuum model (\ref{cont}) that lives on
the {\it bonds} of a square lattice \cite{heiko}. The structure of the
potential is depicted in Figure \ref{hamfig}. We have valleys of effective
depth $-\Delta$, and bulk disorder from a probability distribution
$P(\epsilon_{ij})$ for the bonds of the lattice model.  $P$ is taken to be
either a flat, bounded distribution so that $\left< \epsilon \right> =
\epsilon_{max}/2$, or an unbounded exponential distribution with $\left<
  \epsilon \right> = \epsilon$, the decay constant of the distribution.  Each
bond can only be occupied by a segment of a single line (hard core
interactions) and the occupation of each bond $(ij)$ costs a particular amount
of energy $e_{ij} = \epsilon_{ij} +\Delta_{ij}$ where the $\Delta_{ij}$'s are
zero inside the potential wells and constant elsewhere. This reproduces also
the elastic energy, since all bonds cost some positive energy.  The lattice
Hamiltonian then reads
\begin{equation}
H({\bf x})=\sum_{(ij)} e_{ij}\cdot x_{ij}\;,
\label{hamilflux}
\end{equation}
where $\sum_{(ij)}$ is a sum over all {\it bonds} $(ij)$ joining site $i$ and
$j$ of a $d$-dimensional, e.g.\ rectangular ($L^{d-1}\times H$) lattice, with
open boundary conditions (b.c.) in all space directions. The FL enter the
system via the plane $z=0$ and leave it via the plane $z=H$, see Fig.
\ref{hamfig}. The FL configuration is defined by the set of variables $x_{ij}$
that can take on the values $x_{ij}=1$ (if a line is passing bond $(ij)$ and
$x_{ij}=0$). For the configuration to form {\it lines} on each site of the
lattice the configuration ${\bf x}$ has to be divergence free (i.e.\ 
$\nabla\cdot{\bf x}=0$, where $\nabla\cdot$ denotes the lattice divergence).
All top-sites are attached to an extra site, via energetically neutral arcs,
$e=0$: the source with divergence $\nabla\cdot{\bf x}=+N$, where $N$ is the
number of FLs, and all bottom sites to the target $t$ with $\nabla\cdot{\bf
  x}=-N$.
Finding the minimum energy configuration of $N$ lines now becomes
equivalent to a {\it minimum cost flow problem} with the energy
function (\ref{hamilflux}) and the mass balance constraints expressed
in the site divergences. This task can be solved exactly in
polynomial time (complexity ${\cal O}(N\cdot L^{d-1}H)$
by applying the successive-shortest-path algorithm which is described
in detail in \cite{heiko,flows,Alavaetal}.

\begin{figure}
\epsfxsize=\columnwidth\epsfbox{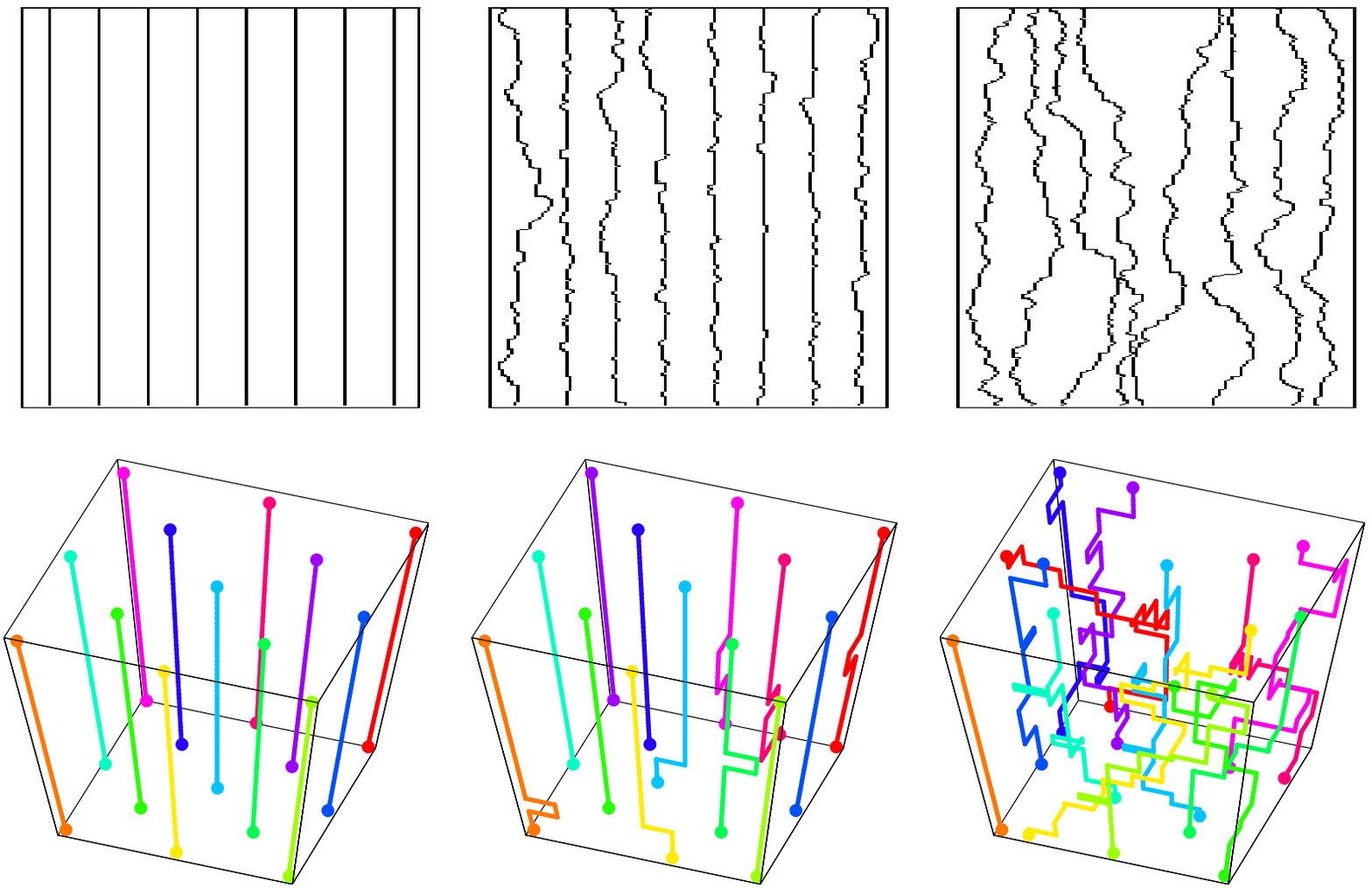}
\caption{
 Optimal ground state configurations in $2D$ (top) and $3D$
 (bottom) for different point disorder strengths $q$, increasing
 from left to right. In the flat phase (left) the FL are trapped
 completely inside the potential valleys.}
 \label{gs}
\end{figure}

Figure \ref{gs} demonstrates with a series of snapshots the geometry
involved in the calculations, and the typical behavior with increasing
$g$ in 2$d$ and 3$d$. In both cases the lines are pinned to
the energetically favorable valleys for small $q$, and finally for
large $q$ a cross-over to a rough state takes place. In 3$d$ one can
observe that the lines wander almost freely in such conditions.  
%

We discriminate between the different regions in the phase diagram by
looking at the behavior of the average transverse fluctuation or {\it
roughness} $w$ of the lines:
\begin{equation}
w(L,H)=\left[\frac{1}{N}\sum_{i=1}^N \frac{1}{H}\int_0^H dz\; 
\Bigl({\bf r}_i(z)-\overline{\bf r}_i\Bigr)^2\right]_{\rm av}\;,
\end{equation}
where $\overline{\bf r}_i=H^{-1}\int_0^H dz\,{\bf r}_i(z)$ and
$[\ldots]_{\rm av}$ denotes the disdorder average. By studying very large
longitudinal system sizes  $H\ge10^4$ we are able to extract
the saturation roughness $w(L)=\lim_{H\to\infty}w(L,H)$ for a finite system of
transverse size $L$. Note that we have chosen open b.c.: the transverse
fluctuations cannot get larger than the system size. Other quatities of
interest are the size $l_{\parallel}$ of the longitudinal excursions (the
average distance between the locations at which a line leaves a valley 
and returns to it); and
the total number of potential valleys $PV$ that a line visits between 
its entry and terminal point in the limit $H\to\infty$.

\begin{figure}[t]
\epsfxsize=\columnwidth\epsfbox{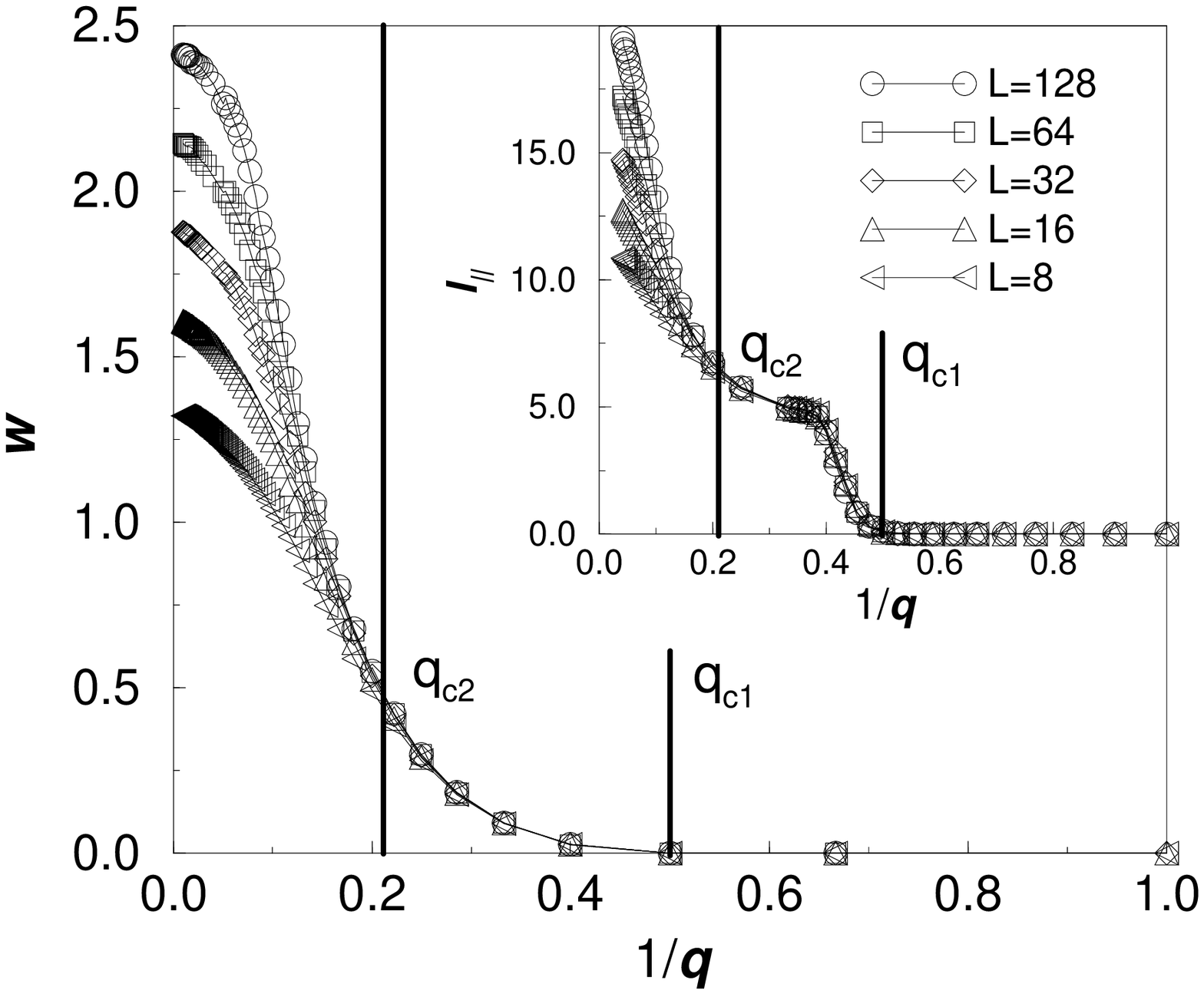}
\caption{
  Roughness $w$ in $2D$ as a function of
  disorder strength $q$ for bounded disorder. $q_{c1}$ and
  $q_{c2}$ are shown. In the flat phase $w=0$, whereas $w>0$ for
  $q>q_{c1}$. No transversal system size dependence is observered
  in the weakly fluktuating phase. The inset shows $l_{\parallel}$.
  Each data point is averaged over $n=20$ ($L=128$) up to $n=600$
  ($L=8$) disorder configurations, $a=4$.
  \label{rough2d}
}
\end{figure}

\begin{figure}[ht]
\epsfxsize=\columnwidth\epsfbox{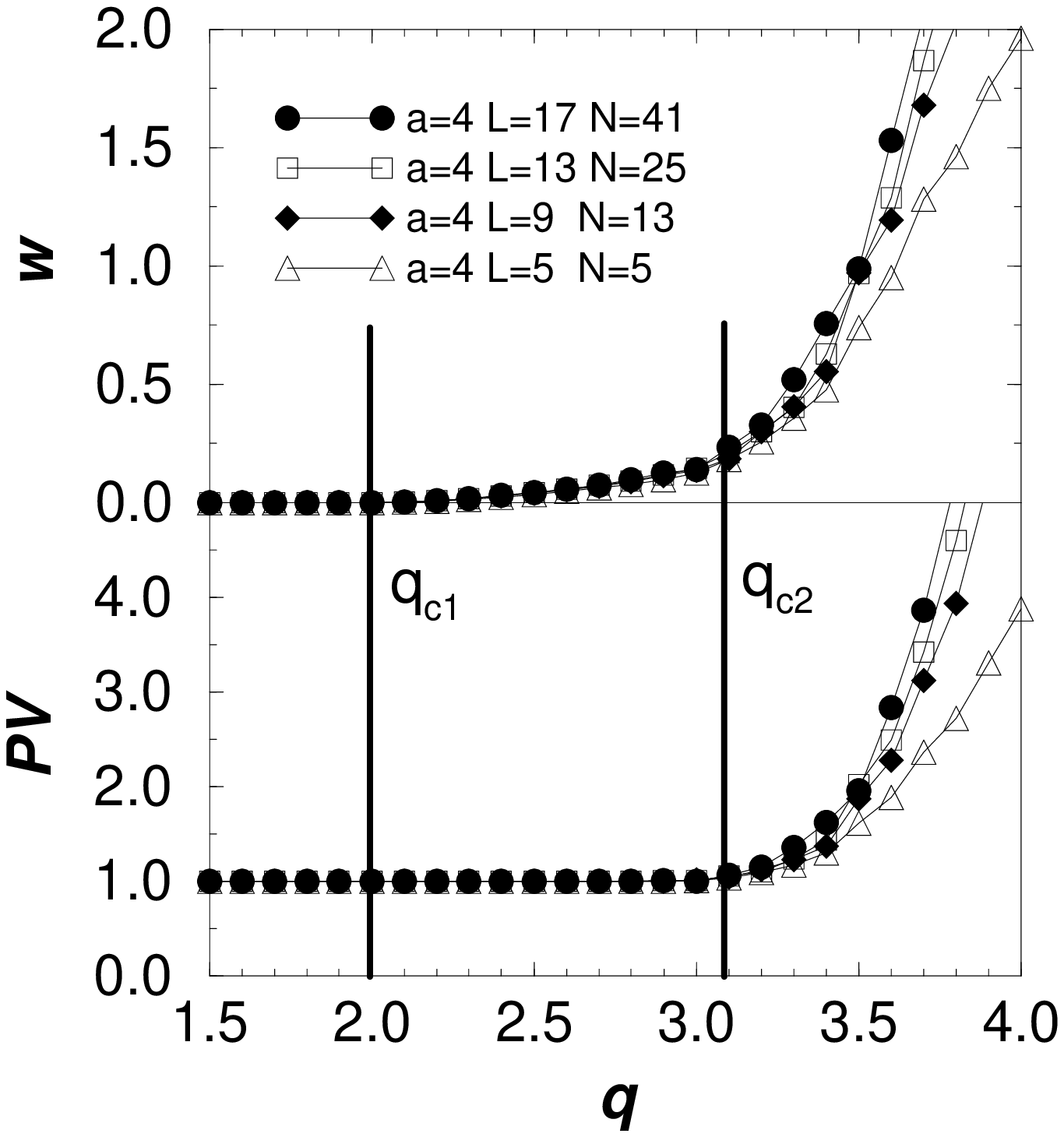}
\caption{
  (Top) The same as in Fig. \ref{rough2d}, now in 3d. (Bottom) The FL
  start to {\em jump} from one potential valley to the next at
  $q_{c2}$ so that the number of potential valleys each line visits
  changes from $pv=1$ to $pv>1$ at this threshold (bottom). $n=6$
  ($L=128$) up to $n=100$ ($L=5$), $a=4$.
  \label{roughET}
}
\end{figure}

\begin{figure}[t]
\epsfxsize=\columnwidth\epsfbox{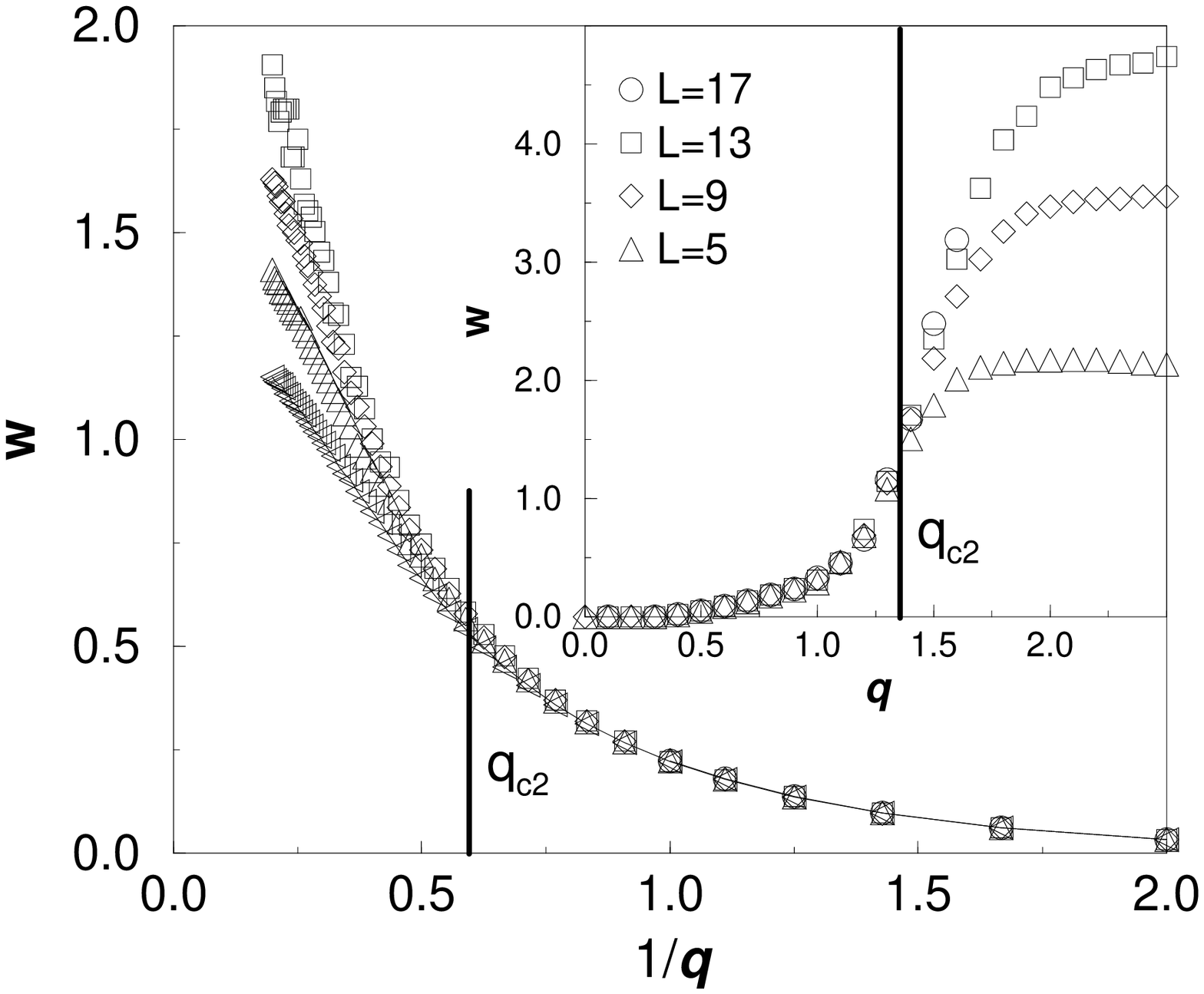}
\caption{
  Same as Fig. ~\ref{roughET} for unbounded
  disorder. The inset shows $3D$. Though hardly visible from this data $q_{c1}$
  vanishes. But we still find a finite value of $q_{c2}$, below which
  the roughness shows no $L$-dependence both, in $2D$ and $3D$. $n=10$
  up to $n=200$, both in $2D$ and $3D$.} 
  \label{roughunb}
\end{figure}

In Fig. \ref{rough2d} we show our data for the roughness $w$ and
$l_{\parallel}$ as a function of $1/q$ in 2d, in Fig. \ref{roughET} $w$ and
$PV$ as a function of $q$ in 3d; both for bounded disorder. In Fig.\ 
\ref{roughunb} we show our data for $w$ for unbounded disorder. The picture
that emerges is the following.
In the {\it flat region} we have $w(L)=0$, $l_{\parallel}=0$ and $PV=1$, i.e.\ 
the lines lie completely in the potential valleys. This region $q<q_{c1}$
exist only for {\it bounded} disorder. For the uniform distribution no
energetically favourable transverse fluctuation can exist as long as
$q<\Delta$. That $q_{c1}>1$ follows from the fact that we are at full
occupancy, $N = N_V$ where $N_V$ is the number of valleys, for $q \le q_{c_1}
\sim 2$ the groundstate consists always of $N$ straight lines regardless of
dimension. For {\it unbounded} disorder this flat region does not exist, since
the probability for a sequence of high-energy bonds in the valleys that pushes
the lines out of it is always positive. In the {\it weakly fluctuating region}
for $q_{c_1} \leq q \leq q_{c_2}$ the lines roughen locally.
Here one has $w>0$ and $l_\parallel>0$, independent of the systems size $L$, 
and  $PV=1$. The transverse fluctuations of flux lines are bounded
by the average line distance or valley separation $a$.
The central feature is that lines fluctuate individually,
so that a columnar defect competes with point disorder. Both in 2D and in 
3D a strong columnar pin strictly
localizes the line \cite{pin} reducing the line-to-line interaction to
zero.

With increasing $q$ the transverse fluctuations increase, and when their
lengthscale $l_\perp$ becomes comparable to the inter-line distance $a$ the
physics changes. When individual lines can jump from
one valley to a neighboring valley, i.e.\ exactly when $PV>1$, a collective
rearrangement of the whole line ensemble becomes possible and manifests itself
in a system size dependence of the saturation roughness (see Fig.\ 
\ref{rough2d} for 2d and \ref{roughET} for 3d). Thus the {\it rough phase} is
characterized by a dependence of $w$ and $l_\parallel$ on the transverse
system size and by $PV>1$. In particular the latter criterion facilitates the
numerical determination of the location of the roughness transition $q_{c_2}$.
We find that the critical disorder strength $q_{c_2}(a)$ increases
monotonically with $a$.

In the limit $q\to\infty$ the collective behavior of the transverse
fluctions of the lines crosses over to that of a line array in the
absence of the periodic potential, though the scaling of displacement
correlations and the system roughness is slightly decreased compared
to the case without the potential because of the lengthscale of the
pinned parts of the lines (the remaining tendency to localize in the
valleys).  For an elastic medium model this would imply that the
correlations should depend in a logarithmic fashion on distance. In
2$d$, in particular, it should be $w\sim\ln L$
\cite{sos_ana,sos_num,natter2} and our data indicate that this
scaling is indeed obtained above $q_{c_2}$.

\begin{figure}[t]
\epsfxsize=\columnwidth\epsfbox{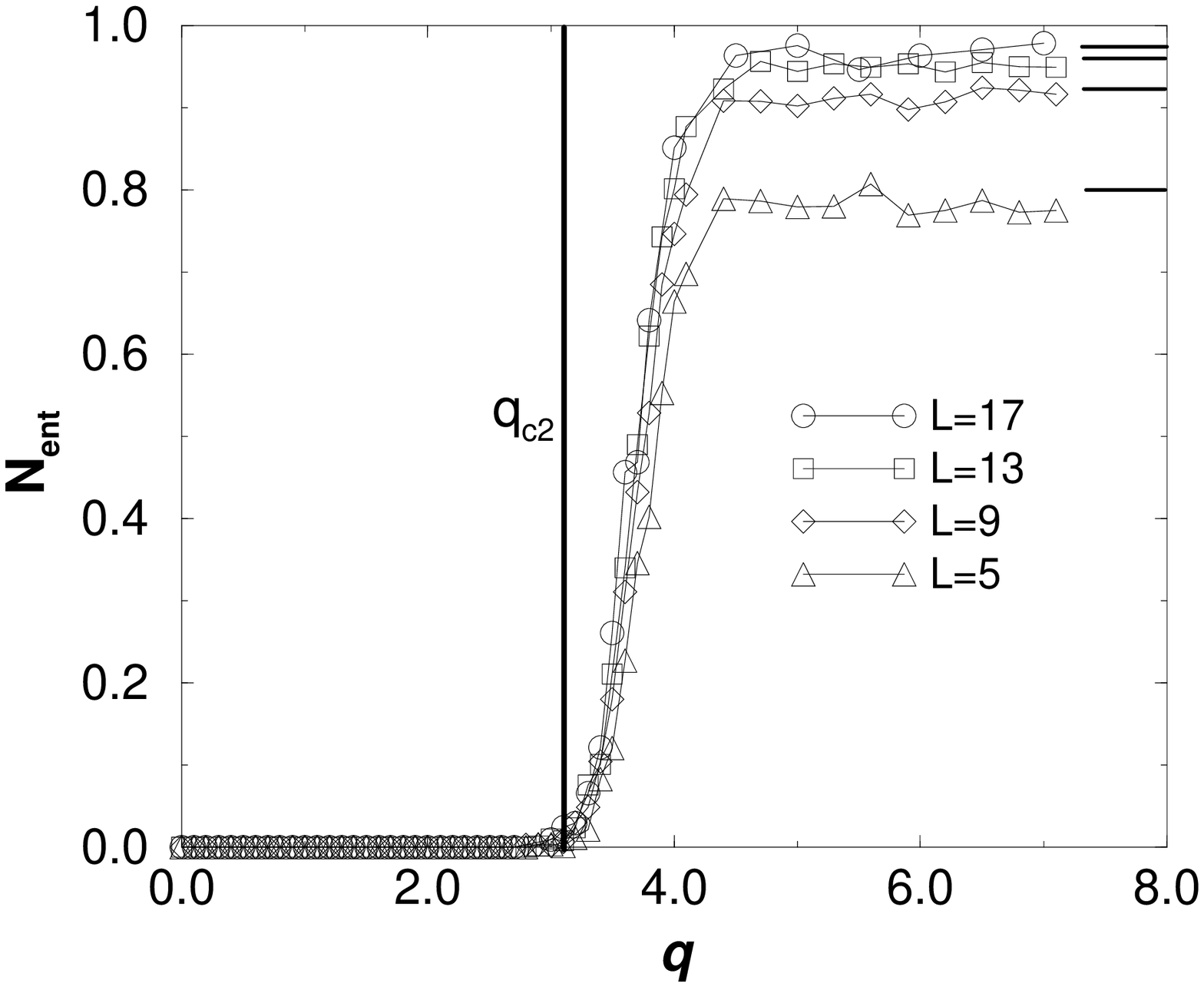}
\caption{
  Number of entangled lines $N_{ent}$ normalized by the 
  total number of lines. The horrizontal lines indicate the
  analytically expected saturation values for random permutations
  of $N$ lines.
  \label{Nent}
}
\end{figure}

A fundamental difference exists between the real-space geometry
of two-dimensional and three-dimensional systems.  In 3D and 
for small line densities, the lines
can wander freely since there is no hard-core expulsion in contrast to
2D. This implies that in the thermodynamic limit, the line ensemble
becomes {\it entangled}. We characterize this by computing the
fraction of lines $N_{ent}/N$ that, when forced to start and end in an
arbitrary one of all the potential minima, enter and exit in different
valleys. In the limits $H\rightarrow \infty$, $L\rightarrow \infty$
$N_{ent} (q)/N$ develops a first-order jump at the roughening
transition from zero to unity. 
$N_{ent} = N$ implies that all the lines are entangled, though in a
finite system the saturation value of $N_{ent}/N$ is 
limited by the finite probability that at least one out of
$N$ integers ends up at exactly at the same position in a
random permutation. This finite size feature can be seen
in the data presented in Fig.~\ref{Nent}.

Finally we relate our findings to periodic elastic media with
point disorder \cite{emig}. These are supposed to represent a coarse
grained description of flux line arrays such as (\ref{cont}) and
predict the {\it absence} of a roughening transition in 2d contrary to
our results. The reason for this discrepancy is that such
models operate in a different limit of many lines
between the potential valleys and interactions that renormalize to an
elastic constant for the line displacements \cite{natter2}. Simple
scaling considerations show \cite{emig} that this model in $d$=1+1 is
always rough, essentially due to the fact that the energy for terrace
excitations of length scale $L$ in the disorder free case
($\lambda=0$) has no bulk contributions. On the other hand, in the
model (\ref{cont}) we have considered here excitations, e.g.\ steps of
individual lines from one potential valley to the next, do have a bulk
contribution, due to the full occupancy of the narrow potential
valleys. Therefore we expect a roughening transition for (\ref{cont})
at a non-vanishing disorder strength, as confirmed by our numerical
results in $d$=2,3

To conclude, we have analyzed with the aid of exact combinatorial
optimization methods ensembles of elastic lines
(or directed polymers or FLs) in the presence of a confining
periodic potential and competing random point disorder. The main
finding is a transition between 'rough' and 'flat regimes,
in both two and three dimensions, at a finite potential strength.
It arises since even rare
fluctuations are not able to induce line-line interactions when
the filling factor of the system is at unity. Introducing 
dislocations (in the form of missing lines) or
extra lines is likely to complicate the physics, as in the
rough substrate model. In the rough phase the physics
is characterized by correlations that increase with system size. 
In three dimensions we find an entangled
phase, in which the lines form a topologically complicated 
geometric configuration. It needs to be characterized properly,
and the effect on the glassy physics should be studied.

This work has been supported by the Academy
of Finland and the German Academic Exchange Service (DAAD)
within a common exchange project, and separately, by
the A. of F.'s Centre of Excellence Programme.

\end{document}